\documentclass{IEEEtran}
\usepackage{cite}
\usepackage{psfrag}
\usepackage{latexsym, amsmath, subfigure, color, amsfonts, amssymb,graphicx}
\usepackage{algorithm}
\usepackage{algorithmic}
\usepackage{epstopdf}

\newtheorem{theorem}{Theorem}[section]

\newtheorem{corollary}[theorem]{Corollary}

\newtheorem{remark}[theorem]{Remark}

\newtheorem{definition}[theorem]{Definition}

\newcommand{\R}{{\rm  I\kern-2pt R}}

\title{Consensus-based Privacy-preserving \\Data Aggregation}

\author{Jianping {He}$^{1}$, Lin Cai$^{2}$, Peng Cheng$^{3}$, Jianping Pan$^{2}$ and Ling Shi$^{4}$
\thanks{$1$: Dept. of Automation, Shanghai Jiao Tong University, and Key Laboratory of System Control and Information Processing, Ministry of Education of China, Shanghai, China {\tt\small jphe@sjtu.edu.cn}}
\thanks{$2$: Dept. of Electrical \& Computer Engineering, University of Victoria, BC, Canada  {\tt\small cai@ece.uvic.ca};  {\tt\small pan@uvic.ca}}
\thanks{$3$: State Key Lab of Industrial
Control Technology, Zhejiang University,  China
{\tt\small pcheng@iipc.zju.edu.cn}}%
\thanks{$4$: Dept. of Electric and Computer Engineering, Hong Kong University of Science and Technology, Clear Water Bay, Kowloon, Hong Kong {\tt\small eesling@ust.hk}}
\vspace{-24pt}}

\begin{document}

\maketitle

\begin{abstract}
Privacy-preserving data aggregation in ad hoc networks is a challenging problem, considering the distributed communication and control requirement, dynamic network topology, unreliable communication links, etc.  Different from the widely used cryptographic approaches, in this paper, we address this challenging problem by exploiting the distributed consensus technique. We first propose a secure consensus-based data aggregation (SCDA) algorithm that guarantees an accurate sum aggregation while preserving the privacy of sensitive data.  Then,  we prove that the proposed algorithm converges accurately and is $(\epsilon, \sigma)$-data-privacy, and the mathematical relationship between $\epsilon$ and $\sigma$ is provided.  Extensive simulations have shown that the proposed algorithm has high accuracy and low complexity, and they are robust against network dynamics.
\end{abstract}

%\begin{IEEEkeywords}
%Data Aggregation, Average Consensus, Privacy Preservation, Ad Hoc Networks
%\end{IEEEkeywords}

\section{Introduction}

%what is the problem and why important, privacy preservation, accurate data aggregation
Privacy-preserving data aggregation (DA) has attracted great attention with many applications in wireless sensor networks, smart metering systems, cloud computing, etc. \cite{jung13infocom, 07heinfocom, 08hemilcom, ozdemir09cn, shi10infocom, groat11infocom,erkin13ispm, saber07}. {We consider the applications in distributed networked systems, where data aggregation can be carried out using consensus algorithms \cite{saber07}.} Typical scenarios include the wireless sensor networks where sensors are deployed randomly in an area to monitor the environment, and the sensing data will be aggregated and polled by a remote monitor; or in a smart metering system where the smart meters collect real-time electricity usage and the aggregated usage in an area will be used by the utility company to adjust power supply and enable appropriate demand control.  However, these data are often privacy-sensitive~\cite{erkin13ispm}.   How to ensure accurate data aggregation while preserving privacy is an essential and challenging issue, especially in ad hoc networks.

%what are the challenges
The ad hoc mode has both pros and cons that should be considered in the design of accurate and privacy-preserving DA. It is well known that in ad hoc networks, centralized algorithm design or optimization solutions are difficult or too costly to implement. Thus, without relying on a centralized controller, an ad hoc network does not suffer from the single-node failure problem and becomes more robust against node failure and link dynamics. On the other hand, without a central trusted authority, it is concerned that some nodes may be compromised or attacked, resulting in the meltdown of the whole network. In addition, dynamic network topology, limited node computing capacity, higher rates of communication errors and losses, and severe delay variations all make privacy-preserving DA more challenging in ad hoc networks.
Although privacy-preserving DA has been heavily investigated, existing solutions are typically based on various cryptography techniques, requiring either secure communication channels, pre-established shared keys, a trusted authority, or the combination of them.

Consensus is an important distributed computing method, which has gained much attention in automatic control and signal processing areas \cite{Olshevsky11, tsp13, huang12, alex14, boyd06tit, cai11tac, carli10auto, ren07acc}, and has been widely used in various networking areas, e.g., time synchronization in sensor networks \cite{lucaautomatica11, he14tac}. Note that an average consensus algorithm can help each node to obtain the average value of all nodes' states in a distributed way, which is a building block of the distributed aggregation algorithm designed in this paper. Recently,  Mo and Murray in \cite{yilin15tac} addressed the privacy-preserving average consensus problem, and they designed a novel Privacy Preservation Average Consensus (PPAC) algorithm to solve the problem. Using PPAC, the privacy-preserving and accurate DA can be achieved in the mean-square sense, while it is more desirable and more challenging to guarantee the privacy and accuracy in a deterministic manner.

%our approach, its novelty and advantage
To meet the above challenges of DA in ad hoc networks, in this work, we investigate the possibility of not relying on cryptography tools. %Instead of maintaining a fixed network topology which might be difficult in ad hoc networks, we utilize the dynamic topology of an overlay network constructed and updated in a distributed way as a means to preserve privacy (\textcolor[rgb]{1.00,0.50,0.25}{How is that? The robustness of the algorithm to topology changes and link failures is shown in simulations, but there is no utilization of dynamic topology to preserve privacy.}).
To enable fully distributed additive data aggregation, we first analyze the conditions on the added noise in the consensus algorithms,  which can guarantee that an average consensus can be achieved deterministically. Then, based on the given conditions, we design a secure consensus-based data aggregation (SCDA) algorithm that can achieve $(\epsilon, \sigma)$-data-privacy and high accuracy in obtaining the sum and the average. Given the accuracy of the aggregation, our solution can be applied to other types of aggregation such as product, variance and other high-order statistics.  %the main contributions of this paper

The main contributions and approaches of this work are summarized as follows. First, we exploited an average consensus algorithm to solve the privacy-preserving data aggregation (DA) problem in ad hoc networks. %This aggregation achieves the privacy preservation property through the designed noise adding process.
We derived a sufficient condition and a necessary condition of the noises added to the consensus process, under which an accurate aggregation is achieved. Based on the sufficient condition, a distributed SCDA algorithm is designed without using any trusted authority, so that the aggregator can obtain the aggregated results from any participating nodes. Second, we proved the convergence of the SCDA. To quantify the degree of the privacy protection, we introduced a novel privacy definition, named $(\epsilon, \sigma)$-data-privacy, which means that the probability that each node can infer its neighbor nodes' initial states in an $\epsilon$ interval is no larger than $\sigma$.  We also proved that SCDA provides $(\epsilon, \sigma)$-data-privacy, and the relationship between $\epsilon$ and $\sigma$ has been derived.

%organization of the paper
The remainder of the paper is organized as follows.  System model and problem formulation are presented in Section~\ref{sec:Prof}.  %Section \ref{sec:mainresults} is the problem simplification. Then,
SCDA is proposed and analyzed in Sections \ref{sec:SCDA}.  Simulation evaluation is presented in Section \ref{sec:simulation}, followed by concluding remarks and further research issues in Section~\ref{sec:conclusions}.

\section{System Model and Problem Formulation}\label{sec:Prof}

\subsection{System Model}
We consider an ad hoc network where nodes are self-organized into clusters (using an existing clustering algorithm~\cite{band03infocom}). We focus on a {\rm connected} cluster with $n$ nodes.  The data from the nodes in the cluster are aggregated, while each individual's data should not be revealed to any other node (including the aggregator) or eavesdropper.  The aggregator can poll any node in the cluster to acquire the aggregated data.

Two nodes can select each other as neighbors to exchange data with a logical link (a single-hop or multi-hop communication
path) between them.   Thus, an underlying network can be constructed. It should be noted that since a logical link can be a multi-hop communication path, the underlying network may not be equivalent to the communication network.  {The application of logical link is to hide the topology information from privacy attackers, and thus it can enhance privacy protection. For example, even an eavesdropper can eavesdrop all one-hop neighbors' information of node $i$, it cannot know which part of the information is used in the state update of node $i$.} The underlying network is modeled as an undirected graph, $G = (V, E)$, where $V$ is the set of nodes and $E$ is the set of logical links (edges) between nodes. Let $N_i$ be the neighbor set of node $i$, where $j\in N_i$ iff $(j, i)\in E$ (neighboring nodes are connected by  logical links).  Note that the logical links are negotiated in a distributed way, and thus node $i$ knows its neighbor set $N_i$, but does not know the full topology of the underlying network.

Let $\mathbf{N}^+$ be the set of positive integers. Define the infinite norm as $\parallel {x} \parallel_\infty=\max \{|x_i|\}$,  which is the maximum absolute value of all the elements of vector ${x}$. We use $\hat{[\circ]}$  to denote an estimation of  $[\circ]$.

\subsection{Problem Formulation}
%At time instant $t_0$, we need to obtain the aggregated data.
Denote the privacy-sensitive data of each node as ${x}_i(0)$, which is also called the initial state of node $i$.
In this paper, we consider how to obtain the additive aggregation, i.e., $\sum_{i=1}^n {x}_i(0)$.
The main design objectives are listed below.
First, the aggregation should be obtained in a distributed manner, without the knowledge of the whole network topology. Second, the computation and  communication cost should be minimized. Lastly,
each node's initial state should not be known to others (including its neighbors, the aggregator, and eavesdroppers) to preserve privacy, while the aggregation should be accurate. %Third, if malfunctioning, selfish, dishonest nodes exist in the system, a distributed safeguard mechanism is needed to fast detect the suspicious behaviours and bound the error in the aggregation due to any undetectable dishonest behaviours.

To achieve the above objectives, we choose to devise the solution based on average consensus which is a well-known distributed algorithm.
Given the number of nodes ($n$),  the sum is easily obtained by multiplying the average of the initial states by $n$. \footnote{Using average consensus, we can obtain the average of $\log x_i$, $(x_i)^k$ (for $k=2,3...$) to calculate the product, variance, and other statistics.}

In a nutshell, distributed average consensus computes the average of the initial data by local information exchanges among neighbors (in the underlying network).
The state of each node is updated iteratively by taking a weighted sum of its current state and those of its neighbors.  If the weights are carefully chosen, the states of all nodes will converge to their average after a number of iterations. To preserve privacy, each state being sent to the neighbors will be added with a noise.
Denote by ${x}_i(k)$ the state of node $i$ at iteration $k$. The information being sent out at $k$-th iteration is designed as
\begin{equation}\label{xiadd}
{x}_i^+(k)={x}_i(k)+{\theta}_i(k), i\in V,
 \end{equation}
where ${\theta}_i$ is the noise for privacy preservation.

In each iteration, the state is updated as follows.
\begin{align}\label{xkjiak}
&{x}_i(k+1)=w_{ii}{x}_i^+(k)+\sum_{j\in N_i} w_{ij} {x}_j^+(k)\nonumber\\
&=w_{ii}({x}_i(k)+{\theta}_i(k))+\sum_{j\in N_i} w_{ij} ({x}_j(k)+{\theta}_j(k))
 \end{align}
for $i\in V$, where $w_{ij}$s are the weights. Here, ${\theta}_i(k)$ may not be necessary, while it is included to simplify the mathematical expression in both the formulation and  proof.

To ensure that average consensus is achieved by the consensus algorithm and that the weights can be obtained in a distributed manner, we
use  Metropolis weights \cite{sboyd05}, given by
\begin{align} \label{mwij}
w_{ij} &= \left\{\begin{aligned}& {  (1+\max\{d_i, d_j\})^{-1} }, &
j\in N_i,
\\
&1-\sum_{l \in N_i} w_{il}, & i=j, \\
&0, & \textrm{otherwise},\end{aligned} \right.
\end{align}
where $d_i$ and $d_j$ are the number of neighbors of node $i$ and $j$ in $G$, respectively. For a connected graph, a matrix with Metropolis weights is  doubly stochastic.
%, which guarantees that an average consensus can be achieved when the nodes exchange correct data (without added noise) for updating~\cite{sboyd05}. %The convergence with noisy data is a key issue that will be studied in the following section.

Putting in the matrix form, we have
\begin{align} \label{iterationdynamic}
&{x}(k+1)=W({x}(k)+{\theta}(k)),
 \end{align}
where ${x}, {\theta} \in R^{n}, W\in R^{n\times n}$ satisfying ${x}=[{x}_1, {x}_2,  ..., {x}_n]^T$ and
${\theta}=[{\theta}_1, {\theta}_2,  ..., {\theta}_n]^T$, and $W$ is the matrix with Metropolis weights as its elements.

%Then, from the definition of Metropolis weights and the elementary matrix transformation, we simplify (\ref{iterationdynamic}) to
%\begin{align} \label{iterationdynamic1}
%\left[\begin{matrix}
%{x}^1(k+1)\\{x}^2(k+1)\\
%...\\
%{x}^m(k+1)
%\end{matrix}\right]&=\begin{bmatrix}
%   W^1 & 0&..&0\\
%    0& W^2 & ..&0\\
%     .. & ..&..&..\\
% 0& 0&..& W^m
%\end{bmatrix}\nonumber \\&\times
%\left(\left[\begin{matrix}
%{x}^1(k)\\{x}^2(k)\\
%...\\
%{x}^m(k)
%\end{matrix}\right]+
%\left[\begin{matrix}
% {\theta}^1(k)\\ {\theta}^2(k)\\
%...\\
% {\theta}^m(k)
%\end{matrix}\right]\right),
% \end{align}
%where ${x}^c$ and ${\theta}^c \in R^{n_c\times l}$ are the states and noise vector of the nodes in cluster $c$, respectively, and $W^c\in R^{n_c\times n_c}$ is the Metropolis matrix, for $c=1,2,...,m$.
%Note that the Metropolis matrix of a connected graph is doubly stochastic \cite{sboyd05}, which means that each $W^c$ is doubly stochastic. Then, one sees that the dynamics in different clusters are separated and similar with each other, i.e.,
%\begin{align} \label{iterationdynamic2}
%&{x}^c(k+1)=W^c({x}^c(k)+{\theta}^c(k)), c=1,2,...,m.
% \end{align}

Define the average state as $\bar{{x}}={1\over n} \sum_{i\in V} {x}_i(0)$.
The problem is to design the noise process ${\theta}(k)$ such that
\begin{align} \label{simplifyobjective}
&\lim\limits_{k\rightarrow \infty} {x}_i(k)=\bar{{x}}, i\in V.
 \end{align}
%for $c=1,2,...,m$, with privacy preservation of each node.

Using the Metropolis weights,  $W$  is doubly stochastic and the average consensus can be easily guaranteed when ${\theta}(k)=0$ for all $k$ \cite{Olshevsky11, tsp13}; however, non-zero noise is necessary to preserve privacy.
If the aggregation can tolerate some discrepancy, we have more freedom to design the noise process ${\theta}(k)$. For example, we can choose ${\theta}(k)$ to be mutually independent with an exponentially decaying co-variance matrix \cite{huang12}. However, to achieve the exact average consensus, the added  ${\theta}(k)$ has to ensure that the consensus result will not be affected and the privacy can be guaranteed, which implies that ${\theta}(k)$ must be carefully designed and correlated. {In \cite{yilin15tac} PPAC was designed to guarantee the privacy and the exact average consensus, by adding and subtracting Gaussian and zero-sum noises to the consensus process. {It is proved that PPAC has a mean-square convergence rate, i.e.,  an exact average consensus can be guaranteed by PPAC in the mean-square sense.} However,  what are the general conditions on the added noise that can guarantee the privacy and the exact average consensus is still an open issue. In the following, we conduct the theoretical analysis and design the algorithm to solve this problem. }
%To simplify the statement,  we denote $W^c=W$, ${x}^c(k)={x}(k)$, and ${\theta}^c(k)={\theta}(k)$ in the following parts of this paper.

%\input{simplification}
%\input{scda}
\section{Private and Accurate Data Aggregation}\label{sec:SCDA}

{In this section, we first analyze the sufficient conditions and the necessary conditions on the added noise process such that a deterministic average consensus can be achieved.  Then, based on the obtained conditions, we propose the SCDA algorithm and analyze its performance in terms of convergence, aggregation accuracy, privacy, and implementation complexity.}

\subsection{Algorithm Design}

%Before given SCDA algorithm,
{We first present a theorem, which provides a sufficient condition of deterministic average consensus and a theoretical support for our algorithm design.}
\begin{theorem}\label{theorem1}
Considering the linear dynamic system (\ref{iterationdynamic}), if  the added noise vectors are bounded, i.e., $\| {\theta}(k)\|_\infty\leq \alpha \rho^k$ for some $\alpha>0$
and $ \rho\in[0, 1)$,  and the sum of all added noises satisfies $\sum_{k=0}^{\infty}\sum_{i=1}^n{\theta}_i(k)=0$,  then we have
\begin{align} \nonumber
&\lim\limits_{k\rightarrow \infty} {x}_i(k)=\bar{{x}}, i\in V.
 \end{align}
Meanwhile, $\sum_{k=0}^{\infty}\sum_{i=1}^n{\theta}_i(k)=0$ is a necessary condition.
\end{theorem}

The proof of Theorem \ref{theorem1} is given in the Appendix \ref{ap_theorem1}, {where the proof of the convergence can be referred to Theorem 3 of \cite{lucaautomatica11}}.
Based on this theorem,  if the noise process ${\theta}(k)$ satisfies the two conditions that $\| {\theta}(k)\|_\infty\leq \alpha \rho^k$, i.e., exponentially decaying, and {$\sum_{k=0}^{\infty}\sum_{i=1}^n{\theta}_i(k)=0$, i.e., zero-sum, the goals of
%(\ref{objec1}) and (\ref{objec2})
accurate and fast aggregation
can be achieved.  The exponentially decaying condition can ensure the convergence of the algorithm. The zero-sum condition ensures that the achieved consensus is an exact average consensus, which guarantees a fully accurate aggregation. Hence, Theorem \ref{theorem1} provides general conditions on the added noise which guarantees that an average consensus can be achieved deterministically.
% We thus have the following  corollary.
%\begin{corollary}\label{corollary1}
%Considering linear dynamic system (\ref{iterationdynamic}), if  $\| {\theta}(k)\|_\infty\leq \alpha \rho^k$ for some $\alpha>0$ and $ \rho\in[0, 1)$, then
%$\lim\limits_{k\rightarrow \infty} {x}_i(k)=C_c$  exponentially fast for $i\in V_c$ and $c=1,2,...,m$, where $C_c=[C_{c1}, C_{c2}, C_{c3}]$ is a constant vector.
%\end{corollary}
%The zero-sum condition ensures that the achieved consensus is an exact average consensus, which guarantees a fully accurate aggregation.
%Therefore, if the adding a noise process with zero-sum and exponentially decaying bound  in (\ref{iterationdynamic2}), an average consensus can be achieved with privacy preservation.
Furthermore, from the proof of Theorem \ref{theorem1}, we have the following corollaries.
\begin{corollary}\label{corollary2}
Consider the linear dynamic system (\ref{iterationdynamic}). If  there are $h$ sub-sequences ${\theta}(\ell+kh)$ of noise process ${\theta}(j)$ and each sub-sequence satisfies $\| {\theta}(\ell+kh)\|_\infty\leq \alpha \rho^k$ for some $\alpha>0$ and $ \rho\in[0, 1)$, and the noise process ${\theta}(\ell)$  satisfies the zero-sum condition, i.e., $\sum_{\ell=0}^{\infty}\sum_{i=1}^n{\theta}_i(\ell)=0$, then
$\lim\limits_{k\rightarrow \infty} {x}_i(k)=\bar{{x}}$ for $i\in V$,  %and $c=1,2,...,m$,
where $\ell=0,1,...,h-1$.
\end{corollary}
%{\begin{corollary}\label{corollary3}
%Consider the linear dynamic system (\ref{iterationdynamic}). If  $\lim\limits_{k\rightarrow \infty} {x}_i(k)=\bar{{x}}$ holds for $\forall i\in V$, we have
%\[\sum_{k=0}^{\infty}\sum_{i=1}^n{\theta}_i(k)=0,\]
%and $\lim_{k\rightarrow \infty}{\theta}(k)=0$ or $\lim_{k\rightarrow \infty}W{\theta}(k)=0$.
%\end{corollary}}

Based on Corollary \ref{corollary2}, each node can randomly divide the noise adding process into several sub-sequences, such that the correlation between any pair of adjacent added noises is not clear to the other nodes.

\begin{algorithm}
\caption{: \emph{SCDA Algorithm}} \label{SCDA}
{\small{
\begin{algorithmic}[1]
\STATE Select each element in ${\theta}_i(0)$ randomly  from $[-\frac{\alpha}{2} \rho, \frac{\alpha}{2} \rho]$.
%c: theta is a vector, so I change the above.
\STATE Let ${x}_i^+(0) = {x}_i(0) +{\theta}_i(0)$ and transmit ${x}_i^+(0)$  to its neighbor nodes.
\STATE Set $\delta_i(0)={\theta}_i(0)$.
\STATE Set $k=1$.
\WHILE{$k < $ Max\_Iteration\_Number} {
%c: are there any convergence condition that we can use to exit from the While loop?
\STATE Update ${x}_i(k)$ with  (\ref{iterationdynamic}) based on ${x}_i^+(k-1)$ and ${x}_j^+(k-1)$ received from all neighbor nodes ($\forall j\in N_i$).

\STATE Select each element of $\delta_i(k)$ randomly or autonomously from $[-\frac{\alpha}{2} \rho^{k+1}, \frac{\alpha}{2} \rho^{k+1}]$, i.e.,
\begin{equation}\label{deltaconstra}
{ |\delta_i(k)| \leq \frac{\alpha}{2} \rho^{k+1}, k\geq 1.}
\end{equation}

\STATE Set  ${\theta}_i(k)$ according to
\begin{equation}\label{thetaconst}
{\theta}_i(k)=
\delta_i(k)-\delta_i(k-1).
\end{equation}

\STATE  Set ${x}_i^+(k)$ using (\ref{xiadd}), and then transmit ${x}_i^+(k)$ to its neighbor nodes.
\STATE $k=k+1$.
}
\ENDWHILE

\end{algorithmic} } }
\end{algorithm}

We further design the SCDA algorithm for node $i$ in Algorithm 1. The Max\_Iteration\_Number in step $5$ is given initially. According to our simulation, we can simply let Max\_Iteration \_Number equal $n^2$, which is sufficiently large to guarantee an accurate aggregation. We can also let each node terminate the iteration when it finds all its neighbors' states are sufficiently close to its own state,  e.g., $|{x}_i(k)-{x}_j(k)|\leq \varepsilon$ for $\forall j\in N_i$ and a given small $\varepsilon$. SCDA is a fully distributed algorithm. Only the neighbor set $N_i$ is the input of each node $i$, and after sufficient iterations ($k\geq n^2$), all nodes' updated states could be the output of SCDA. Based on the output, the aggregator can easily achieve the goal of DA. In addition, we can also use the same approach given in \cite{yilin15tac}  to prove that  SCDA also converges at least in a mean-square sense.

%For the SCDA algorithm, the steps $1$ and $2$ are the initialization of the algorithm, which gives the initial setting of the parameters used for each node. Step $3$ means that each node will update its state according to an average iteration process (\ref{iterationdynamic}). The key design is step $4$, which describes how to choose the noise process such that the two conditions in Theorem \ref{theorem1} can be satisfied. The last step $5$ is the noise adding process, which is to preserve the privacy of nodes.

\subsection{Convergence and Accuracy of SCDA}
The following theorem gives the convergence and accuracy of SCDA, and its proof is given in the appendix.
%it is  proved that the SCDA algorithm achieves an average consensus with an exponential convergence rate.

\begin{theorem}\label{theorem2}
Using the SCDA algorithm, we have $\lim\limits_{k\rightarrow \infty} {x}_i(k)=\bar{{x}}$ for $\forall i\in V$, i.e., an average is achieved.
%and
%\begin{equation} \nonumber
%\lim\limits_{k\rightarrow \infty} {x}_i(k)=\bar{{x}}_c, i\in V_c
%\end{equation}
%for $c=1,2,...,m$,
\end{theorem}

For each cluster, every node will achieve an average consensus using the SCDA algorithm, i.e., the aggregator can obtain the average state $\bar{{x}}$ from any node  after the algorithm converges. Then, the sum can be obtained from using $n\bar{{x}}$, resulting in an accurate sum aggregation.
%Hence, a fully accurate aggregation is achieved by SCDA algorithm.

\begin{remark}
It follows from Theorem \ref{theorem2} that  for SCDA, there exists $k_0>0$ such that $V(x(k))<\varepsilon$ holds for $\forall k\geq k_0$ and $\varepsilon>0$, where $V(x(k))=\max({x}(k))-\min ({x}(k))$. However, this is not true for PPAC. The reason is that
\begin{align*}
V({x}(k+1))&=V(W({x}(k)+{\theta}(k))) \\&\geq |V(W{\theta}(k))-V(W{x}(k))|,
\end{align*}
where $\Pr\{V(W{\theta}(k))\geq M\} >0$
holds for any $M>0$ since  $f_{\theta_i(k)}(y)>0$ holds for $\forall\theta_i(k) \in \theta(k)$ and $y\in R$. Thus, one infers that $V({x}(k)) <\varepsilon$ cannot be guaranteed by PPAC for any given $k>0$ with probability 1.
\end{remark}

 Note that the proof of  Theorem \ref{theorem2} only used the properties of a doubly stochastic matrix and the results given in Theorem \ref{theorem1}.  SCDA can also be adopted to solve the privacy of the asynchronous gossip consensus algorithms which also have the doubly stochastic matrixes in the algorithm dynamic functions, e.g., \cite{boyd06tit, cai11tac, carli10auto}. However, considering the privacy of more complicated consensus algorithms, e.g., second-order consensus, e.g., \cite{ren07acc}, it is a more challenging and open problem.

\begin{remark}
 {With SCDA, a higher accuracy of DA requires more iterations and an exact DA needs a sufficiently large number of iterations. It should be noticed that the larger communication delays will decelerate the convergence speed of SCDA. Hence, when the delays are not negligible, there is a tradeoff between convergence speed and DA accuracy, and we will discuss how to accelerate the convergence speed of SCDA at the end of this section. }
\end{remark}

\subsection{Privacy of SCDA}\label{security}
For SCDA, node $i$ only transmits the information sequence ${x}_i^+(k)$, $k=0, 1, ...$, to its neighbors. For each message ${x}_i^+(k)$, there is a noise component ${\theta}_i(k)$ added to ${x}_i(k)$. Hence, any neighbor node cannot know the exact value of ${x}_i(0)$ based on the received information sequence from node $i$. Meanwhile, note that when $k\geq 1$, ${x}_i(k)^+$ is an updated state which may be quite different from the initial state ${x}_i(0)$, since each update is an average process among all the information received from its neighbor nodes' states.
{Define for $\forall j\in N_i$, the information set which is available for node $i$ at iteration $k$ as follows,
\begin{align*}
  \mathcal{I}_{ij}(k)=&\{x_i(0), x_i^+(0), ...,
  x_i(k), x_i^+(k); \\ &  x_{\ell}^+(0), ..., x_{\ell}^+(k),\forall \ell\in  N_i\},
\end{align*}
where all the message of node $i$ and the message output of its neighbors are included in $ \mathcal{I}_{ij}(k)$, and let $\mathcal{I}_{ij}(\infty)=\lim_{k\rightarrow \infty}\mathcal{I}_{ij}(k)$.}
Suppose that node $i$ cannot listen to all the neighbors' information of node $j$. This assumption can be guaranteed in the underlying network construction with $N_j\nsubseteq N_i$, and it has been proved to be necessary in \cite{yilin15tac}. The added noises are assumed to be unknown to each node $i$, and the initial states of nodes are independent with each other.

{Note that if node $i$ does not have any prior information of $x_j(0)$ and no additional information is available for estimation, then it is unlikely to make an accurate estimation on $x_j(0)$ with a high probability.  That is,  we cannot make an accurate estimation directly if we do not have any information about the initial state of a node.} Hence, when node $i$ directly estimates node $j$'s initial state without using any prior or side information, denoted by $\hat{x}_j^0(0)$, it is reasonable to assume
\begin{equation}\label{intiassum}
\Pr\{ \hat{x}_j^0(0) \in  [x_j(0)-\epsilon, x_j(0)+\epsilon]\}\ll \sigma,
\end{equation}
where $\epsilon$ and $\sigma$ are two given small positive constants, and $\sigma=\max_{\nu\in[-\frac{\alpha}{2} \rho, \frac{\alpha}{2} \rho]} \int_{\nu-\epsilon}^{\nu+\epsilon} f_{\theta_j(0)} (y) \text{d} y$. This assumption can be extended to the case that side information may be available. For instance, if it is known that the state $x_j(0)$ is belong to the interval $[-M, M]$ with equal probability,  we have
\[\Pr\{ \hat{x}_j^0(0) \in  [x_j(0)-\epsilon, x_j(0)+\epsilon]\}={\epsilon\over M}.\]
In this case, (8) still holds if there exists $[v-\epsilon, v+\epsilon]$ such that $f_{\theta_j(0)} (y)\gg {1\over M}$ for $\forall y\in[v-\epsilon, v+\epsilon]$.

Under SCDA, the broadcast information of node $j$, i.e.,  $x_j^+(0), x_j^+(1), ..., x_j^+(k)\in \mathcal{I}_{ij}(k)$,  is available to node $i$ to infer/estimate the initial value of neighbor node $j$. Note that  the information output, $x_j^+(k)$,  equals the weighted sum of the received information in the previous round plus a noise. Based on these information output, node $i$ will take the probability over the space of all noises $\{\theta_j(k)\}_{k=0}^\infty$ (where the space is denoted by $\Theta$) under the condition that $\mathcal{I}_{ij}(\infty)$ is known,  to estimate the values of the added noises. Then,  using the difference between each information output and the estimated noises, we have $\hat{x}_j(0)= {x}_j^+({k})-\hat{\theta}_j^k$, where $\hat{\theta}_j^k$ is the estimation of random noise $\theta_j^k$ ($\theta_j^k={x}_j^+({k})-{x}_j(0)$). Using this estimation, we have $|\hat{x}_j(0)-x_j(0)|=|\hat{\theta}_i^k-\theta_i^k|$,  and
 \begin{align}  \label{deprivacyi}
 \Pr\left\{ |\hat{x}_i(0)-x_i(0)|\leq \epsilon\right\}&=\Pr\left\{ |\hat{\theta}_i^k-\theta_i^k|\leq \epsilon\right\}.
 %\\&=,
\end{align}

To evaluate the privacy of SCDA, we give the definition of $(\epsilon, \sigma)$-data-privacy as follows.
\begin{definition}
A distributed algorithm provides $(\epsilon, \sigma)$-data-privacy, if, with information set $\mathcal{I}_{ij}(\infty)$, the probability that each node $i$ can successfully estimate its neighbor node $j$'s initial value $x_j(0)$ in a given interval $[x_j(0)-\epsilon, x_j(0)+\epsilon]$ is no larger than $\sigma$, i.e.,
 \begin{equation}\label{deprivacy}
\sigma=\max_{\hat{\theta}_i^k\in \Theta, k\geq 0}\Pr\{|\hat{\theta}_i^k-\theta_i^k|\leq \epsilon\}.
\end{equation}
 \end{definition}
 In the above definition,  $\epsilon$ indicates the estimation accuracy  and $\sigma$
expresses the privacy cost. Given the estimation accuracy $\epsilon$, a smaller value of $\sigma$ offers a stronger privacy guarantee.

\begin{remark} {For noise-adding privacy preserving solutions, no matter what type of noise distribution is used, there is a chance that an estimated value of the original data is close to the real data, but such a probability cannot be directly measured by differential privacy or the privacy metrics based on mutual information or Fisher information (e.g., given an estimation accuracy, the disclosed probability of initial states cannot be measured by the existing privacy metrics directly).  Hence, it motivates us to introduce $(\epsilon, \sigma)$-data-privacy, which is defined as the probability of $\epsilon$-accurate estimate (the difference of an estimation and the original data is within $\epsilon$) is no larger than $\sigma$ (the disclosure probability). This definition reveals the relationship between the privacy and the estimation accuracy.  Therefore, the propose privacy definition links the disclosure probability and the estimation accuracy directly, which is meaningful to quantify the data privacy  in the applications of consensus. } %In addition, for the proposed privacy definition, we assume that the privacy attacker has the local information only for estimation, but for the existing ones, e.g., differential privacy, the attacker may have and apply all the information output of nodes for estimation.}
%The proposed $(\epsilon, \sigma)$-data-privacy reveals the relationship between the maximum data disclosure probability ($\sigma$) under a given estimation accuracy
%range ($\epsilon$).
%On the other hand, differential  privacy ensures that an adversary cannot infer whether the data  is associated with a user or not with a high probability.
%In addition, for $(\epsilon, \sigma)$-data-privacy, the privacy attacker can use the information output of the node and the state of itself, i.e., the local information, to estimate the original data. But for the most existing privacy metrics, the privacy attacker can use the global information in addition to the initial data of the node to reveal user's identity.
\end{remark}

Next, we prove that  SCDA provides  $(\epsilon, \sigma)$-data-privacy, and a theorem is stated as follows.
\begin{theorem} \label{theoremprivacy}
SCDA algorithm is $(\epsilon, \sigma)$-data-private, and the relationship between $\epsilon$ and $\sigma$ satisfies
\begin{equation}
\sigma=\max_{\nu\in[-\frac{\alpha}{2} \rho, \frac{\alpha}{2} \rho]} \int_{\nu-\epsilon}^{\nu+\epsilon} f_{\theta_j(0)} (y) \text{d} y,
\end{equation}
and $\lim_{\epsilon\rightarrow 0}\sigma=0$, where $f_{\theta_j(0)} (y)$ is the probability density function (PDF) of $\theta_j(0)$.
\end{theorem}

\begin{remark}
It should be noted that Theorem \ref{theoremprivacy} is obtained under the assumption that node $i$ cannot listen to all the neighbors' information of node $j$. If this assumption is relaxed and node $i$ has the knowledge of $N_j$, then at any time $k\geq 1$, node $i$ can exactly calculate the value of  $\theta_j(k)$ through the following equation,
\[\theta_j(k)={x}_j(k)-[w_{jj}{x}_j(k-1) +\sum_{l\in N_j} w_{jl} {x}_l^{+}(k-1)],\]
where all the expressions on the right-hand side are known to node $i$. Hence, over the time, node $i$
can calculate all of $\theta_j(k), ..., \theta_j(1)$. Then, using the zero-sum property of the noise,  node $i$ can calculate $\theta_j(0)$ by
$\theta_j(0)=-\sum_{k=1}^\infty \theta_j(k)$.
Therefore, node $i$ knows the value of $x_j(0)$ through $x_j(0) = x_j^+(0)-\theta_j(0)$, i.e.,  $x_j(0)$ is released. This result  is consistent
with Theorem 4 in \cite{yilin15tac}, which proved that the disclosed space of a node with $m$ neighbors is of dimension
$m + 1$.
\end{remark}

\subsection{Complexity of SCDA}\label{complexity}
%Last, we analyze the computation and communication complexity of the SCDA algorithm.
Since each node just calculates a weighted average at each iteration, SCDA has very low computation complexity, in  $O(n)$.  According to our simulation results, when the underlying network is well connected (e.g., the diameter of the graph is much smaller than $n$), the consensus can be reached in $O(n)$ iterations. Note that the number of hops is confined to the diameter of the cluster,  we can also let nodes select logical neighbors within a small number of hops (e.g., $1$ to $3$). Thus, the communication cost is in $O(kn^2)$, where $k$ is the number of iterations which is typically smaller than $n$ for large $n$. %Note that in our modeling, a well connectivity topology of a cluster is not difficult to be guaranteed. For example, we can let each node connect with more nodes in each cluster to enhance the connectively of network topology. And,
We can further divide the network into more clusters to accelerate the convergence rate, while as a trade-off the aggregator needs to poll more nodes. The latest consensus algorithm proposed in \cite{alex14} can guarantee that an average consensus is achieved in a few iterations, or nearly linear time. It thus can be applied to guarantee an ultrafast average consensus, which further reduce the communication cost of SCDA.

\section{Conclusions and Further Discussions}\label{sec:conclusions}
In this paper, we have investigated the privacy-preserving data aggregation problem in ad hoc networks using the average consensus approach. We have proposed the SCDA algorithm to solve the problem. SCDA is simple to implement and can ensure private and accurate aggregation.  {SCDA does not rely on a centralized controller or a trusted aggregator, and it can be implemented in a distributed manner and robust against the network dynamics.}  Simulation results have shown that the proposed algorithm has fast convergence rate and high accuracy, and they are robust against network dynamics.

There are still many open issues worth further investigation. In this paper, the underlying network should be a connected, undirected graph. To ensure connectivity, a spanning tree connecting all the nodes in the cluster can be built and the links in the spanning tree should be included in the underlying network. %In this work, the connected requirement refers to the jointly connected constraint \cite{sboyd05, saber07}, i.e., temporary node failures or disconnections during the iterations can be tolerable, as demonstrated in our simulation results.
How to deal with permanent node failures needs further investigation. The undirected graph requires bi-directional communications. In case bi-directional logical link cannot be maintained, novel consensus solutions need to be used which are much more complicated. % Second, different from the SCDA case, to deal with dishonest nodes, in E-SCDA, the aggregator should have the knowledge of the topology of the underlying network, and how to relax this requirement is an interesting while challenging problem. A possible direction is to design an incentive mechanism such that all nodes are willing to be honest so as to achieve an accurate privacy-preserving aggregation at a lower cost. Third, the E-SCDA approach can help to constrain the dishonest nodes, assuming that the selected nodes will not generate false reports and there is no intrusive node in the system. How to secure the system to deal with intrusive nodes remains an open issue.
We have proved that SCDA can converge exponentially, while the exact convergence speed remains an open issue.
Overall, using distributed consensus can be a promising alternative to the heavily investigated privacy-preserving approaches using cryptography techniques in ad hoc networks and other distributed systems. It is also possible to combine these two powerful tools to further enhance privacy and security, or make a good tradeoff between computation and communication complexity, which beckons further research.

\appendices

%\vspace{15pt}

\section{The proof of Theorem \ref{theorem1}} \label{ap_theorem1}

\begin{proof}
{First, we prove that each ${x}_i(k)$ is bounded by some constant $M$ for $i\in V$. Since $W$ is doubly stochastic, we have $\parallel W \parallel_{\infty}=1$. Hence,
\begin{align} \label{iterationdynamicproof}
&~~\parallel{x}(k+1)\parallel_{\infty}=\parallel W({x}(k)+{\theta}(k)) \parallel_{\infty}
\nonumber\\&
\leq \parallel W \parallel_{\infty}\parallel{x}(k)+{\theta}(k) \parallel_{\infty}
\leq \parallel{x}(k) \parallel_{\infty}+\parallel{\theta}(k) \parallel_{\infty}
\nonumber\\
&\leq \parallel{x}(0) \parallel_{\infty}+\sum_{\ell=0}^k \parallel{\theta}(\ell) \parallel_{\infty}.
\end{align}
Using the condition that  $\| {\theta}(\ell)\|_\infty\leq \alpha \rho^\ell$, it follows
\begin{align} \label{xkcproof}
\parallel{x}(k+1)\parallel_{\infty}
&\leq \parallel{x}(0) \parallel_{\infty}+\sum_{\ell=0}^k \alpha \rho^\ell
\nonumber\\&\leq \parallel{x}(0) \parallel_{\infty}+\frac{\alpha}{1-\rho}=M,
\end{align}
which implies that each ${x}_i(k)$ is bounded by $M$ for all $k$.

Next, we prove the convergence of  (\ref{iterationdynamic}). % Define a function as follows:
%\begin{align}
%V({x}(k))= \max({x}(k))-\min ({x}(k)).
%\end{align}
The function $V({x}(k))$
is nonnegative and has the property that $V({x}(k))=0$ if and only if all the elements of ${x}(k)$ have the same values, i.e., ${x}(k)=C {1}$, where $C$ is a constant and ${1}$ is a vector with all its elements equal to $1$.}

{Note that $W^\ell$ is still a doubly stochastic matrix for $\ell\in {N}^+$, and we have $\lim_{\ell\rightarrow \infty}W^\ell={1\over n}{1}^T{1}$. Since the topology of each cluster is assumed to be connected, we have $W^{n}>0$. Then,  from Lemma $2$ in \cite{lucaautomatica11}, it follows that, for any vector ${y}$,
\begin{align}\label{fact1}
\max\{W^{n}{y}\}-\min\{W^{n}{y}\}\leq (1-\epsilon)(\max\{{y}\}-\min\{{y}\}),
\end{align}
where $\epsilon=\max_{j=1}^{n}\min_{i=1}^{n}(W^{n})_{ij}, \epsilon\in(0, 1)$. Hence, we have
\begin{align}\label{vfunction}
&V({x}(k+n))= \max({x}(k+n))-\min ({x}(k+n)) \nonumber
\\ &\leq  \max(W^{n}{x}(k))-\min (W^{n}{x}(k))\nonumber
\\ &+\sum_{\ell=0}^{n} [\max( W^{\ell}{\theta}(k+n-\ell))-\min ( W^{\ell}{\theta}(k+n-\ell))]
\nonumber\\& \leq (1-\epsilon) V({x}(k))+2\sum_{\ell=0}^{n} \alpha \rho^{k+n-\ell} \nonumber\\& \leq (1-\epsilon) V({x}(k))+{2 \alpha {\rho^{k}(1-\rho^{n+1})\over 1-\rho}},
\end{align}
where we used the fact of (\ref{fact1}). From (\ref{vfunction}), one infers that
\begin{align*}
&V({x}(\ell+h n))\leq (1-\epsilon) V({x}(\ell+(h-1)n))+\hat{\alpha}(\ell) \rho^{(h-1)n}
\nonumber\\
&\leq (1-\epsilon)^2 V({x}(\ell+(h-2)n))\nonumber\\&+\hat{\alpha}(\ell)[ \rho^{(h-1)n}+(1-\epsilon)\rho^{(h-2)n}]
\nonumber\\&\leq (1-\epsilon)^l V({x}(\ell+(h-l)n))\nonumber\\& +\hat{\alpha}(\ell)[\rho^{(h-1)n}+(1-\epsilon)\rho^{(h-2)n}+...+(1-\epsilon)^{l-1}\rho^{(h-l)n}] \nonumber\\
&\leq (1-\epsilon)^h V({x}(\ell))+\hat{\alpha}(\ell)h \max\{\rho^{(h-1)n}, (1-\epsilon)^{(h-1)}\},
\end{align*}
for $\ell=0,1,..., n-, l=3,4,...,h$ and $h\in{N}^+$, {where $\hat{\alpha}(\ell)= 2 \alpha \rho^\ell {(1-\rho^{n+1})\over 1-\rho}$ is a constant}.
{Since $\epsilon\in(0, 1)$ and $\rho\in[0,1)$, $\lim_{h\rightarrow \infty}V({x}(\ell+h n))=0$
for $\ell=0,1,..., n-1$. Clearly, the above equation implies that
$\lim_{k\rightarrow \infty}V({x}(k))=0$,
i.e.,
\begin{align}\label{limitx1}
\lim_{k\rightarrow \infty}\max({x}(k))-\min({x}(k))=0,
\end{align}
{which means that the differences between elements of ${x}(k)$ will converge to zero. Then, we will prove that the sum of each column vector of ${x}(k)$ is a constant,  and thus prove that an average consensus can be achieved.}}

{Define $\sum(\circ)$ as the sum of all elements in $(\circ)$.
Since $W$ is still a doubly stochastic matrix, we have $\sum(W{x}(k))= \sum({x}(k))$. Then, one obtains that
\begin{align}\label{limitx4}
\sum({x}(k))&=\sum(W{x}(k-1)
+W{\theta}(k-1))\nonumber\\&= \sum({x}(0)+\sum_{\ell=0}^{k-1}{\theta}(\ell)).
\end{align}
Taking limiting on both sides of the above equation yields
\begin{align}\label{limitx5}
\lim_{k\rightarrow \infty}\sum({x}(k))&= \sum({x}(0)+\lim_{k\rightarrow \infty}\sum_{\ell=0}^{k-1}{\theta}(\ell))
=\sum({x}(0)),
\end{align}
where we used the condition that $\sum[\sum_{\ell=0}^{\infty}{\theta}(\ell)]=0$. Combining (\ref{limitx1}) and (\ref{limitx5}) yields that
$\lim_{k\rightarrow \infty}{x}(k)=C {1}=\bar{{x}}{1}$,
i.e., an average consensus is achieved.

It notes from (\ref{limitx4}) that
\[\lim_{k\rightarrow \infty}\sum_{i=1}^n{x}_i(k)= \sum_{i=1}^n{x}_i(0)+\lim_{k\rightarrow \infty}\sum_{i=1}^n \sum_{\ell=0}^{k-1}{\theta}_i(\ell).\]
If (\ref{simplifyobjective}) holds, then $\sum_{i=1}^n{x}_i(\infty)=\sum_{i=1}^n{x}_i(0)$.  It thus follows that the zero-sum condition is the necessary condition to achieve an exact average consensus with (\ref{iterationdynamic}).}}
\end{proof}

\section{The proof of Theorem \ref{theorem2}}\label{ap_theorem2}

\begin{proof}
{We just need to prove that the SCDA algorithm can ensure the two conditions in Theorem \ref{theorem1}.}

{First, we prove that the first condition, i.e., $\sum_{k=0}^{\infty}\sum_{i=1}^n{\theta}_i(k)=0$, is ensured by SCDA. From step 1 and 4, one infers that ${\theta}_i(1)+{\theta}_i(0)=\delta_i(1)$ and ${\theta}_i(2)+{\theta}_i(1)+{\theta}_i(0)=\delta_i(2)$ for any $i\in V$.  Then, by mathematical induction,  one obtains that
$\sum_{k=0}^{\infty} {\theta}_i(k)=\lim_{k\rightarrow \infty}\delta_i(k)$.
From (\ref{deltaconstra}), one has that
\[\lim_{k\rightarrow \infty}| \delta_i(k)|\le\lim_{k\rightarrow \infty}|\frac{\alpha}{2} \rho^{k+1}|=0,\]
which implies that $\sum_{k=0}^{\infty} {\theta}_i(k)=0$ for any $i\in V$. Hence, we have $\sum_{k=0}^{\infty}\sum_{i=1}^n{\theta}_i(k)=0$.}

{Next we prove the added noise, ${\theta}(k)$, is exponentially decaying, i.e., $\| {\theta}(k)\|_\infty\leq \alpha \rho^k$. From (\ref{thetaconst}) and (\ref{deltaconstra}), one infers
\begin{align}
|{\theta}_i(k) | &=|\delta_i(k)-\delta_i(k-1) | \leq |\delta_i(k)|+| \delta_i(k-1) | \nonumber\\
&\leq \frac{\alpha}{2} \rho^{k+1} +\frac{\alpha}{2} \rho^{k}\leq {\alpha} \rho^{k}.
\end{align}
Thus, we have $\| {\theta}(k)\|_\infty\leq \alpha \rho^k$.  }
\end{proof}

\section{The proof of Theorem \ref{theoremprivacy}} \label{ap_theoremprivacy}
\begin{proof}
To prove this theorem, we need to prove that at each iteration $k$, the probability that each node $i$ can successfully infer that $x_j(0)\in[x_j(0)-\epsilon, x_j(0)+\epsilon]$ is no larger than $\sigma$ using the information set $\mathcal{I}_{ij}(k)$. In the following, we prove this result for each iteration.

{At time $k=0$, node $i$ can estimate neighbor $j$'s initial value based on $ \mathcal{I}_{ij}(0)$ and use the fact that \begin{align}\label{eq:xj0}
x_j^+(0)=x_j(0)+\theta_j(0),
\end{align}
for estimation. Then, the corresponding estimation is given by
\begin{align}\label{eq:xje0}
x_j^+(0)=\hat{x}_j(0)+\hat{\theta}_j(0).
\end{align}
Then,  we have
\begin{align}
& \Pr\left\{ \hat{x}_j(0)\in [x_j(0)-\epsilon, x_j(0)+\epsilon]  \right\}
=  \Pr\left\{ |\hat{\theta}_j(0)-\theta_j(0)|\leq \epsilon \right\}\nonumber \\
& =\Pr\left\{ \theta_j(0) \in [\hat{\theta}_j(0)-\epsilon, \hat{\theta}_j(0)+\epsilon] \right\}
=\int_{\hat{\theta}_j(0)-\epsilon}^{\hat{\theta}_j(0)+\epsilon} f_{\theta_j(0)} (y) \text{d} y.
\end{align}
Note that  $\hat{\theta}_j(0)$ is an estimation and could be any values in $[-\frac{\alpha}{2} \rho, \frac{\alpha}{2} \rho]$. Hence, we have
\begin{align}\label{eresult0}
&\max_{ \hat{x}_j(0)}\Pr\left\{ \hat{x}_j(0)\in [x_j(0)-\epsilon, x_j(0)+\epsilon] \right\}  \nonumber \\
= & \max_{\nu\in[-\frac{\alpha}{2} \rho, \frac{\alpha}{2} \rho]} \int_{\nu-\epsilon}^{\nu+\epsilon} f_{\theta_j(0)} (y) \text{d} y
\end{align}
%Note that $f_{\theta_j(0)} (\nu)$ is a continuous and bounded function, and thus we can set $\sigma(0)=2 \epsilon \max_{\nu\in[-\frac{\alpha}{2} \rho, \frac{\alpha}{2} \rho]}  f_{\theta_j(0)} (\nu)$ to guarantee $(\epsilon, \sigma(0))$-privacy.
Hence, $(\epsilon, \sigma)$-data-privacy is ensured at time $k=0$ for SCDA.
}

At time $k=1$, node $i$ can estimate $x_j(0)$  based on $\mathcal{I}_{ij}(1)$ and use the fact of both (\ref{eq:xj0}) and
\begin{align} \label{eq:xj1}
&{x_j^+(1)\over w_{jj}}=x_j^+(0)+{1\over w_{jj}} [\sum_{l\in N_j} w_{jl} x_l^+(0)+\theta_j(1) ]\nonumber \\
&=x_j(0)+\theta_j(0)+{1\over w_{jj}} [\sum_{l\in N_j} w_{jl}(x_l(0)+ \theta_l(0))+\theta_j(1) ] .
\end{align}
If using (\ref{eq:xj0}) only, we also have (\ref{eresult0}). Then, we consider the estimation with (\ref{eq:xj1}).
{Let $f_{\theta'_j(1)} (z)$ be the PDF of ${\theta}'_j(1)$, where
\begin{align} {\theta}'_j(1)&=\theta_j(0)+{1\over w_{jj}}[\sum_{l\in N_j} w_{jl} x_l^+(0)+\theta_j(1)] \nonumber \\
&=\theta_j(0)+{1\over w_{jj}}\theta_j(1)+\sum_{l\in N_j} {w_{jl}\over w_{jj}} x_l^+(0) \nonumber \\
&=\theta_j(0)+{1\over w_{jj}}\theta_j(1)+{\theta}^{''}_j(1).
\end{align}
Based on  (\ref{eq:xj1}), one can make estimation, ${x_j^+(1)\over w_{jj}}=\hat{x}_j(0)+\hat{\theta}'_j(1)$.
Let  $\tilde{\theta}_j(1)=\theta_j(0)+{1\over w_{jj}}\theta_j(1)$. Then, we have
\begin{align}\label{eq:pre1}
  &~~\max \Pr\{|\hat{\theta}'_j(1)- \theta'_j(1)|\leq \epsilon\}\nonumber\\ &
\leq \max \Pr\{|\hat{\theta}^{'}_j(1)- \theta^{'}_j(1)|\leq \epsilon | \tilde{\theta}_j(1)\}   \nonumber \\
&= \max \Pr\{|\hat{\theta}^{'}_j(1) - \tilde{\theta}_j(1)- \theta^{''}_j(1)|\leq \epsilon\}  \nonumber \\
&\leq \max \Pr\{|\hat{\theta}^{''}_j(1)- \theta^{''}_j(1)|\leq \epsilon\},
\end{align}
where $\hat{\theta}^{''}_j(1)=\hat{\theta}^{'}_j(1) - \tilde{\theta}_j(1)$ is viewed as an estimation of $\theta^{''}_j(1)$, and we have used the independence between variables $\theta_j(0)+{1\over w_{jj}}\theta_j(1)$ and $\theta^{''}_j(1)$.
Since node $i$ cannot listen to all the neighbors' information of node $j$, there exists at least one independent variable $x_l^+(0)$ in $\sum_{l\in N_j} w_{jl} x_l^+(0)$ that is unknown to node $i$ (i.e., there is no information of $x_l^+(0)$ available to node $i$) to estimate the value of $\theta^{''}_j(1)$.  From (\ref{intiassum}), it follows that
\begin{align} \label{eq:pre1ad}
 & \Pr\{|\hat{\theta}^{''}_j(1)- \theta^{''}_j(1)|\leq \epsilon\}
%=& \int_{\hat{\theta}'_j(1)- \epsilon}^{\hat{\theta}'_j(1)+\epsilon} f_{{\theta}'_j(1)|{{\theta}^{''}_j(1) }} (z) d z\nonumber \\
%=& \int_{\hat{\theta}_j(0)- \epsilon}^{\hat{\theta}_j(0)+\epsilon} f_{{\theta}_j(0)} (z) d z\nonumber \\
 \leq \max_{\nu\in[-\frac{\alpha}{2} \rho, \frac{\alpha}{2} \rho]} \int_{\nu-\epsilon}^{\nu+\epsilon} f_{\theta_j(0)} (y) \text{d} y.
\end{align}
Combining (\ref{eq:pre1}) and (\ref{eq:pre1ad}), we have
\begin{align*}% \label{eq:pre1ad}
 & \Pr\{ \hat{x}_j(0) \in  [x_j(0)-\epsilon, x_j(0)+\epsilon]\} \nonumber \\
 \leq & \max_{\nu\in[-\frac{\alpha}{2} \rho, \frac{\alpha}{2} \rho]} \int_{\nu-\epsilon}^{\nu+\epsilon} f_{\theta_j(0)} (y) \text{d} y.
\end{align*}}
Then,  using  (\ref{eq:xj0}) and  (\ref{eq:xj1}) together for estimation, we have
\begin{align} \label{eq:pre2}
&\Pr\{\hat{x}_j(0) \in [x_j(0)-\epsilon, x_j(0)+\epsilon]\} \nonumber \\
 \leq & \max_{\substack{\nu \in[-\frac{\alpha}{2} \rho,  \frac{\alpha}{2} \rho] \\ \mu \in[b_1, B_1]}}\int_{\nu-\epsilon}^{\nu+\epsilon} \int_{\mu-\epsilon}^{\mu+\epsilon} f_{{\theta}_j(0), {\theta}'_j(1)} (y, z) \text{d} z \text{d} y
 \nonumber \\
 \leq & \max_{\substack{\nu \in[-\frac{\alpha}{2} \rho,  \frac{\alpha}{2} \rho] \\ \mu \in[b_1, B_1]}}\int_{\nu-\epsilon}^{\nu+\epsilon} \int_{\mu-\epsilon}^{\mu+\epsilon} f_{{\theta}'_j(1)|{\theta}_j(0)} (z|y)  f_{{\theta}_j(0)} (y) \text{d} z \text{d} y
 \nonumber \\
 \leq & \max_{\nu \in[-\frac{\alpha}{2} \rho, \frac{\alpha}{2} \rho]}\int_{\nu-\epsilon}^{\nu+\epsilon}   f_{{\theta}_j(0)} (y) \text{d} y,
\end{align}
 where $[b_1, B_1]$ ($B_1-b_1>0$) is an interval including all the possible value of ${\theta}'_j(1)$.
The above result means that if we combine the two facts for estimation, it will not enhance the successful estimation probability. Therefore, at time $k=1$,  we still have (\ref{eresult0}) and $(\epsilon, \sigma)$-data-privacy is still ensured.

{At each iteration $k$,  with similar analysis, there are $k+1$ facts (equations) can be used for estimation. {Based on the  ($k+1$)-th equation, i.e.,
\begin{align} \label{eq:xjk1}
&{x_j^+(k)\over [W^k]_{jj}}={1\over [W^k]_{jj}}\left[ [W^k]_j{x}(0)+\sum_{l=0}^k[W^{k-l}]_j {\theta}(l) \right]
\nonumber \\ &=x_j(0)+\tilde{\theta}_j(k) \nonumber \\ &
+ \left[{[W^k]_j \over [W^k]_{jj}}{x}(0)+\sum_{l=0}^k{[W^{k-l}]_j \over [W^k]_{jj}} {\theta}(l)-x_j(0)-\tilde{\theta}_j(k)\right]
\nonumber \\ &=x_j(0)+ \tilde{\theta}_j(k)+ {\theta}^{''}_j(k)=x_j(0)+ {\theta}'_j(k)
\end{align}
where $[W^k]_j$ denotes the $j$-th row vector of $W^k$, $[W^k]'_j$ is a vector obtained from setting $[W^k]_{jj}=0$ for $[W^k]_j$, and $\tilde{\theta}_j(k)=\sum_{l=0}^k{[W^{k-l}]_{jj}\over [W^k]_{jj}} {\theta}_j(l)$. Then, with the similar analysis of (\ref{eq:pre1}) and (\ref{eq:pre1ad}), we can obtain the following equation,
\begin{align*}
&\Pr\{\hat{x}_j(0) \in [x_j(0)-\epsilon, x_j(0)+\epsilon]\} \nonumber \\
\leq& \Pr\{|\hat{\theta}^{''}_j(k)- \theta^{''}_j(k)|\leq \epsilon\}
 \leq \max_{\nu\in[-\frac{\alpha}{2} \rho, \frac{\alpha}{2} \rho]} \int_{\nu-\epsilon}^{\nu+\epsilon} f_{\theta_j(0)} (y) \text{d} y,
\end{align*}
where we have used the fact that $\hat{\theta}^{''}_j(k)$ contains the independent variables with no information available to node $i$.}
Also, if we combine the equations together, we can prove that the successful estimation probability cannot be increased. That is,   (\ref{eresult0}) holds and $(\epsilon,  \sigma )$-data-privacy is proved at iteration $k$.}

{From the above discussion,  one concludes that  (\ref{eresult0}) holds and $(\epsilon, \sigma)$-data-privacy is guaranteed by SCDA. Meanwhile, note that $ f_{{\theta}_j(0)} (\nu)$ is the PDF function of ${\theta}_j(0)$, it follows that $\lim_{\epsilon\rightarrow 0}\sigma=0$. }
\end{proof}

\end{document}